# IMPACT OF SOFTWARE REQUIREMENT VOLATILITY PATTERN ON PROJECT DYNAMICS: EVIDENCES FROM A CASE STUDY

Rahul Thakurta

Information Systems, Xavier Institute of Management, Bhubaneswar, India

rahul@ximb.ac.in

Subhajit Dasgupta

Business Application Services – JDEdwards practice, Wipro Technologies, India

subhajit.dasgupta@wipro.com

## ABSTRACT

*Requirements are found to change in various ways during the course of a project. This can affect the process in widely different manner and extent. Here we present a case study where-in we investigate the impact of requirement volatility pattern on project performance. The project setting described in the case is emulated on a validated system dynamics model representing the waterfall model. The findings indicate deviations in project outcome from the estimated thereby corroborating to previous findings. The results reinforce the applicability of system dynamics approach to analyze project performance under requirement volatility, which is expected to speed up adoption of the same in organizations and in the process contribute to more project successes.*

## 1. INTRODUCTION

Software developers nowadays have reconciled to the fact that requirements *will* change during the development of software [1, 2]. Such change in requirements during software project lifecycle referred to as requirement volatility has been found to adversely impact project outcomes like effort, time and residual errors [3, 4, 5]. Requirement change has also been observed to take place in different patterns (for example exponential rise (NASA case study: [6]), exponential decay [7], triangular [8], etc) where a pattern indicates the geometrical shape to which the change orders/requests generating during project development can be approximated. For a given amount of requirement change, the results showed disproportional variation in project parameters like effort, schedule, manpower and error generation with the pattern of requirement volatility [9], and the findings were in contradiction to the COCOMO estimates (Constructive Cost Model [10]). However the findings were based on data of a medium-scale project under hiring and in presence of schedule penalty. Given the findings of the study, can we expect similar results in live-project settings? How well does the simulation findings portray and explain the project dynamics in organizations? To facilitate investigation, we adopt a case-study approach here where-in a validated model of software project dynamics is calibrated to the project environment.

The paper is organized in the following sections. The description of the case is provided in the next section. Then we provide the methodology where we present the model that has been used here, and outline the experiment design. The following section presents the study results.

Finally in conclusion we summarize the key findings and also present the future research opportunities.

## 2. CASE EXAMPLE

The case study was conducted at a leading information technology services organization with headquarter in USA, and offices worldwide. The IT service is organized as an onsite/offshore delivery model and uses industry standard frameworks for providing solutions to the business. The projects are executed using project management methodologies like the waterfall, the iterative, and the agile frameworks. The teams caters to the organizational capabilities for performing the various project related activities like planning, scheduling and tracking, review and audit, requirements management, test management, defect and issue management. Detailed data about the process is regularly captured and stored in the software environment.

This particular project is based on the waterfall model and was found to be endangered because of requirement volatility. The project's data consists of estimates of project size, effort, duration and manpower, number and type of the change requests raised and the associated effort, and specific values of parameters needed to synchronize the model with the project environment. The data were collected from available project metrics and based on discussions with project members.

### 2.1. Project overview

The project involved updating the "Advanced Commercial Banking System (ACBS)" of a leading bank based in US. The lending department of the bank used a lower version of the ACBS which needed to be upgraded to version 4.05. The project was estimated to be of medium-sized (< 10,000 Lines of Code (LOC) for which the waterfall process model was chosen as appropriate. The project was initially planned for a one-release cycle of 34 weeks (170 working days) starting from 02-April-2002 (onsite requirements analysis), with the implementation tentatively ending on 30-October-2002. The representatives of the lending department group serves both as the user and the business side contact to the project. The project was executed on the IBM AS400 platform and used $3^{rd}$ party tools for development. The project team comprised of developers, quality assurance engineer, and the project manager. Personnel from support and other areas are involved in the project, but are not treated as members of the development team. The project characteristic is summarized in Table 1.

Table 1. Project characteristics

| Parameter | Value / Description |
| --- | --- |
| Project Name | Advanced commercial Banking System Upgrade |
| Brief Project Description | Commercial Lending department of a leading Bank in US uses ACBS. Needs upgrade from a lower version to ACBS 4.05 |
| Project Type | Conversion / Application Upgrade |
| Development Platform | IBM AS400, LANSA AD CASE tool |
| Programming Language | RPG/400, CL/400 LANSA Rapid Development Maintenance Language |
| Application Type | Banking Application to Handle Commercial Loans |
| Project Life Cycle Model | Waterfall |

Because of some start up delays, the project ultimately started on 01-June-2002, 2 months behind the planned starting date. The initial project manpower was 3 persons. The project spanned 65 calendar weeks (325 working days) ending finally on 31-Aug-2003. The delay was primarily because of issue of change requests during project execution, which introduced difficulties in project management. The project was also holdup for a month to synchronize implementation with other modules, which contributed to the delay. The final delivered project size was approximately 9985 LOC which led to a total expenditure of 2452 man-days of effort on development, quality assurance (QA), rework, and testing activities.

## 2.2. Change Requests

Several change requests were raised by the users during project development. The changes requested were of nominal complexity. Table 2 lists the type of change requests raised, its priority, the start and end dates, the effort expended, and the final status of the change requests.

Table 2. Change requests

| CR No | Priority | Change Type | Status | Actual Estimate | | |
|---|---|---|---|---|---|---|
| | | | | **Start Date** | **End Date** | **Actual Effort** |
| 1 | Urgent | Add | Completed | Sept 2002 | Dec 2002 | 20 man-days |
| 2 | Urgent | Add | Completed | Sept 2002 | Dec 2002 | 27 man-days |
| 3 | Desirable | Add | Completed | Dec 2002 | Dec 2002 | 3 man-days |
| 4 | Desirable | Query | Completed | Feb 2003 | Feb 2003 | 6 man-days |
| 5 | Desirable | Report | Completed | Feb 2003 | Feb 2003 | 3 man-days |
| 6 | Desirable | Modify | Completed | Feb 2003 | Aug 2003 | 28 man-days |
| 7 | Desirable | Report | Completed | March 2003 | March 2003 | 5 man-days |
| 8 | Desirable | Query | Completed | March 2003 | March 2003 | 4 man-days |
| 9 | Desirable | Modify | Completed | March 2003 | March 2003 | 6 man-days |
| 10 | Urgent | Impact analysis | Completed | April 2003 | May 2003 | 21 man-days |

The cumulative total effort expended in these 10 change requests was 123 Person-Days. The changes were raised during a span of 11 months (between September, 2002 and August, 2003 with the exclusion of January in which no new change requests were raised or resolved). All the change requests were resolved successfully and incorporated in the project. This resulted in a further increase in project size by 2414 LOC from the original estimate.

## 3. METHODOLOGY

System dynamics (SD) [11] has been used increasingly in software development [12] to model different problems and conduct what-if type analysis to assist various stakeholders. The basic premise in SD is that system behavior results from interaction among its feedback loops. Model building begins with development of a causal loop diagram that consists of a collection of causal links, each having a certain polarity. A positive (negative) link implies a reinforcing (balancing) relation where a positive change in the cause results in a positive (negative) change in the effect. The causal loop graph can be mapped to a mathematical model consisting of a system of difference equations, which can be simulated under different parametric conditions.

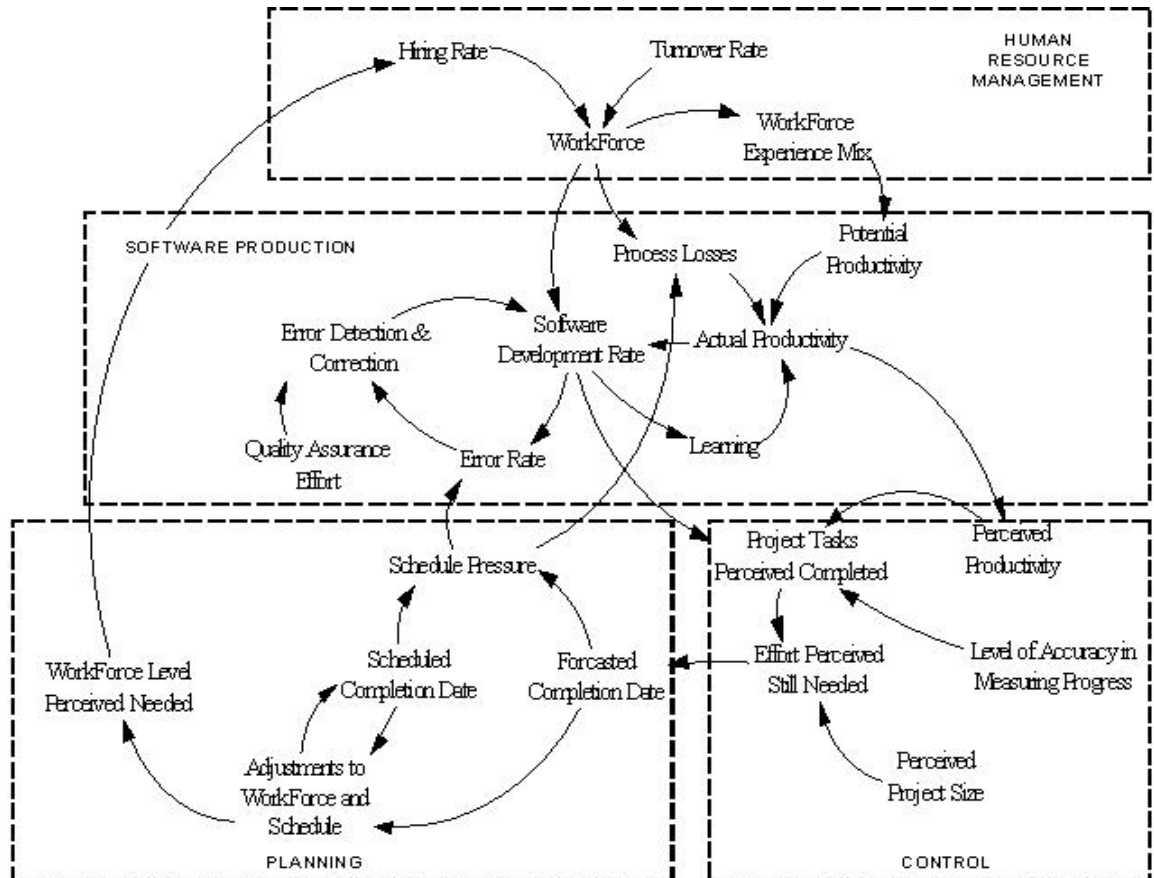
Figure 1. Model causal-loop diagram

Our starting point is Abdel-Hamid's [6] SD model based on the waterfall methodology that integrates all relevant processes of software development. The model causal loop diagram is shown in Figure 1. The arrows in the diagram represent cause-effect relationship e.g. *Schedule Pressure* affects *Error Rate*. *Perceived Project Size* culminates in *Project Tasks Perceived Completed* as *Workforce* at *Actual Productivity* level work on the project at *Software Development Rate*. *Workforce* size changes as a result of *Adjustments to Workforce and Schedule* decision of the management and the resulting *Hiring rate*. *Actual Productivity* of workforce is affected by *Potential Productivity*, *Process Losses* and *Learning*. Rookies in the team affect *Workforce Experience Mix* unfavorably and lower *Actual Productivity*. Increase in *Workforce* size increases *Process Losses* and deteriorates *Actual Productivity*. Increase in *Forecasted Completion Date* increases *Schedule Pressure* and, in turn, increases *Actual Productivity*.

In reality the change in dynamics due to change in *Perceived Project Size* is far more complex because of delays in various cause-effect links. For example, organizations take time to find right people and allocate them to projects. Rookies also take time to get trained and become fully productive. This introduces delay between *Adjustments to Workforce and Schedule* decision and *Workforce*. The increase in *Effort Perceived Still Needed* caused by increase in *Perceived Project Size* thus takes time to affect increase in *Project Tasks Perceived Completed* and subsequent downward adjustment of *Effort Perceived Still Needed*.

The model was simulated in order to investigate the impact of the change order generation pattern based on data of change requests (refer to Table 2) using the commercially available iThink software. Given below are the estimates of some of the key parameters required for simulation

- **Perceived Project Size**

  At the beginning of the project, the project size was estimated at 7572 Lines of Code. There were subsequent additions and modifications because of requirements volatility

- **Estimated Project Effort**

  The initial estimate of project effort from beginning of requirements analysis till the end of big fixing was 780 man-days. Our model excludes requirements analysis & prototyping, implementation & acceptance testing phases, and subsequent maintenance support phases. For this we had to deduct the following:
  - 75 man-days (for requirements analysis & prototyping)
  - 65 man-days (for implementation & acceptance testing phases)
  - 10 man-days (for maintenance support phase)

  Hence the final estimate of project effort for our simulation model came out as 630 man-days

- **Estimated Project Schedule**

  The project was initially estimated to be 34 weeks (170 working days) starting from 02-April-2002. Since our model excludes requirements analysis & prototyping, offshore infrastructure set-up, implementation & acceptance testing and subsequent support phase, we subtracted the following from the estimate:
  - 20 working days (for requirements analysis & prototyping)
  - 5 working days (for offshore infrastructure set-up)
  - 45 working days (for implementation & acceptance testing)
  - 10 working days (for offshore support)

  Thus the effective schedule estimate for our model was arrived at 90 working days (18 weeks)

- **Nominal Potential Productivity**

  This parameter represents the set of productivity determinants that distinguish different development environments, such as availability of software tools, languages used, computer hardware characteristics, and product complexity [6]. This nominal potential productivity remains invariant during the development process of a single project. The nominal potential productivity for this project was estimated based on the actual effort expended on project development. This includes effort expended on development, QA, and rework activities. As stated above, the total effort expended to develop 9985 Lines-of-Code was approximately 2452 man-days (excluding effort spend on requirements gathering, implementation and support). The person day expenditure on QA, rework were not recorded separately. Testing accounted for about 30% of the above calculated effort based on the initial specification. Therefore, the effort expended solely on project development activities accounted for about 70%* 2452 = 1716 man-days.

  Based on the above, the software development productivity came out as 9985/1716 = 5.82 LOC/man-day. In our model, the nominal productivity is the productivity, considering the multiplier due to motivation and communication loss, and multiplier due to project complexity. Multiplier due to project complexity was estimated at 0.75 based on the data provided. Multiplier due to user involvement came out at 0.58. The nominal fraction of man-day on the project was 0.7 and the project on average used 5 full-time personnel. These two led to the derivation of multiplier due to motivation and communication loss as (0.7*(1 - 0.03)) = 0.679. The nominal potential productivity was thus estimated as 5.82/(0.679*0.75*0.58) = 19.7 LOC/man-day.

- **Initial Staffing Level**

The project began with three full time employees (FTE) who were experienced in the project domain.

## 4. RESULTS

The model was run to simulate the project outcome. Here we analyze how the change order generation pattern impacts the project dynamics, and in the process compare the simulation results with the actual behavior. The discussion on the key parameters is provided below

**Change Order Generation Rate**

Figure 2 depicts the pattern of change order generation during project execution based on the simulation results. The actual project values are shown as red squares. The actual values were arrived by assuming that the effort expended on each of the completed change requests were expended uniformly over the prescribed duration. This was then converted into appropriate units (Tasks/Day) which then represented the average rate of change of requirements. The X-axis represents time in working days.

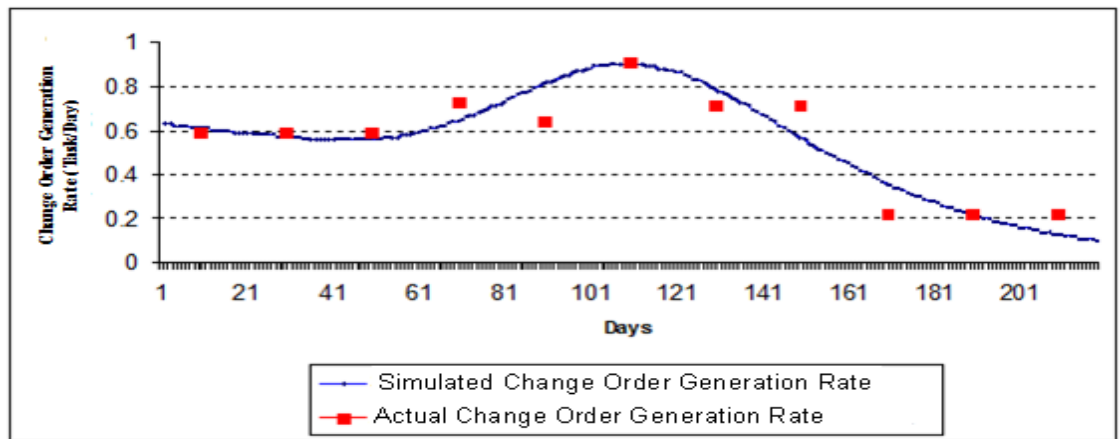

Figure 2. Change order generation rate

Results depict striking similarities in the pattern of change order generation between the actual and the simulated output. The change requests raised during the initial stages of the project were comparatively large and involved long processing duration. In the middle stages a high priority change request was raised which also needed immediate attention. Some small change requests were raised towards the end of the project. Such behavior was also reproduced by the simulation model. The simulation model output indicated a final delivery of precisely 504.3 tasks (9935 LOC) which is very close to the actual result (9985 LOC)

**Total Workforce**

Figure 3 depicts the workforce augmentation pattern in both the simulation model, and the actual project. The red curve indicates the actual project workforce at any point of time. The simulated outcome is provided by the blue curve.

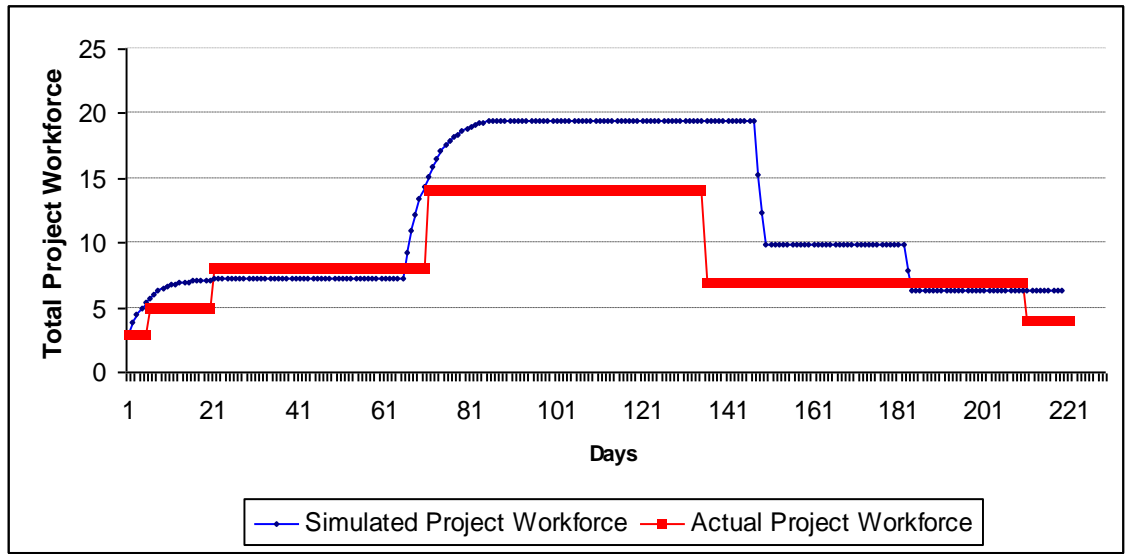

Figure 3. Total workforce

In the real project, the workforce augmentation followed a discrete pattern. Project hiring was based on a workforce allocation matrix developed at the start of the project. Some deviations from the planned matrix did take place because of workforce availability issues, and management decisions.

The simulation model represents workforce augmentation in a continuous manner. The upfront uniform rate of change order generation (Figure 2) led to some initial hiring after which the workforce stabilized. Another phase of hiring was triggered from day 60 onwards driven by an increase in the rate of change order generation (Figure 2). The workforce was gradually released in the last stages when the rate of change order generation also dropped down. This continuous pattern of workforce adjustments led to a higher peak of total workforce compared to the real scenario in which the hiring and release were in discrete intervals. Results indicate the simulated workforce pattern to exceed the actual result. The project workforce was decided in agreement with the business side and was billed accordingly. Hence it was not possible for the project management to change the project workforce at will.

**Software Development Productivity**

Figure 4 plot the simulated software development productivity over time together with actual project results (red squares).

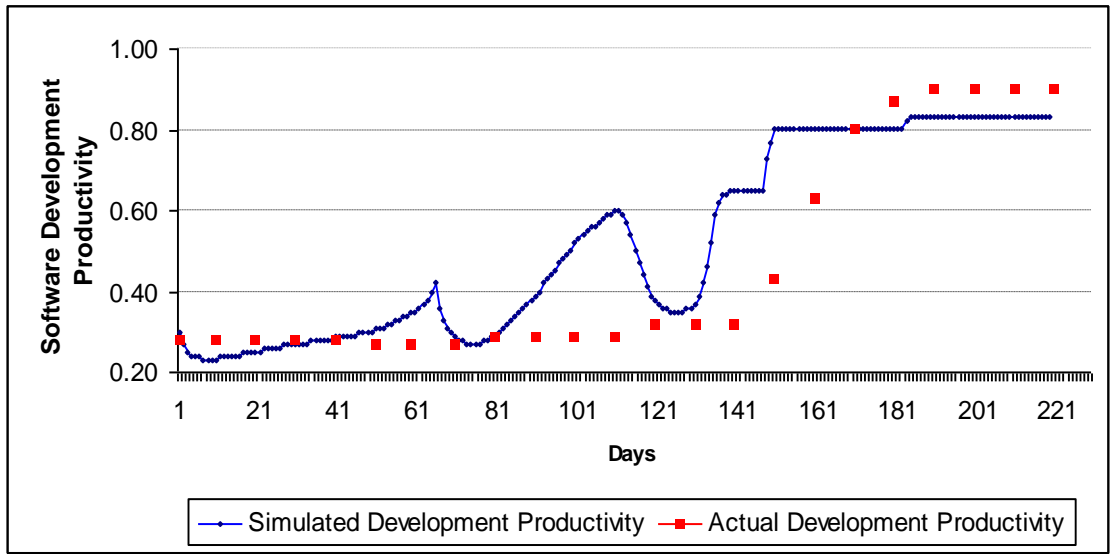

Figure 4. Software development productivity

The variation of productivity in the simulation model can be explained as follows. The workforce stability attained during the initial portion of the project (Figure 3) led to a gradual rise in productivity. Some dip occurs at a later instance triggered by the phase of hiring (Figure 3) and the resultant communication and training overheads. The schedule pressure (not shown) increases towards the later stages of the project, and it causes the productivity to peak which then continues till project completion.

Now under the actual scenario, the productivity data was collected at discrete points, roughly at intervals of two weeks. The productivity was low to start with as a new technology was used in the project with which the project members were not very competent. There were not much observed fluctuations in productivity during the initial stages of the project and the pattern was pretty uniform. Productivity rapidly increased towards the later stages of the project as the workforce became experienced with the technology, and they were also working long hours per day. The pattern of increase closely matched the simulation result.

**Schedule Completion Date**

Figure 5 depicts how the project estimated completion date, measured in terms of number of working days, changed during the project. The actual project values are shown as red squares. The X-axis represents time in working days.

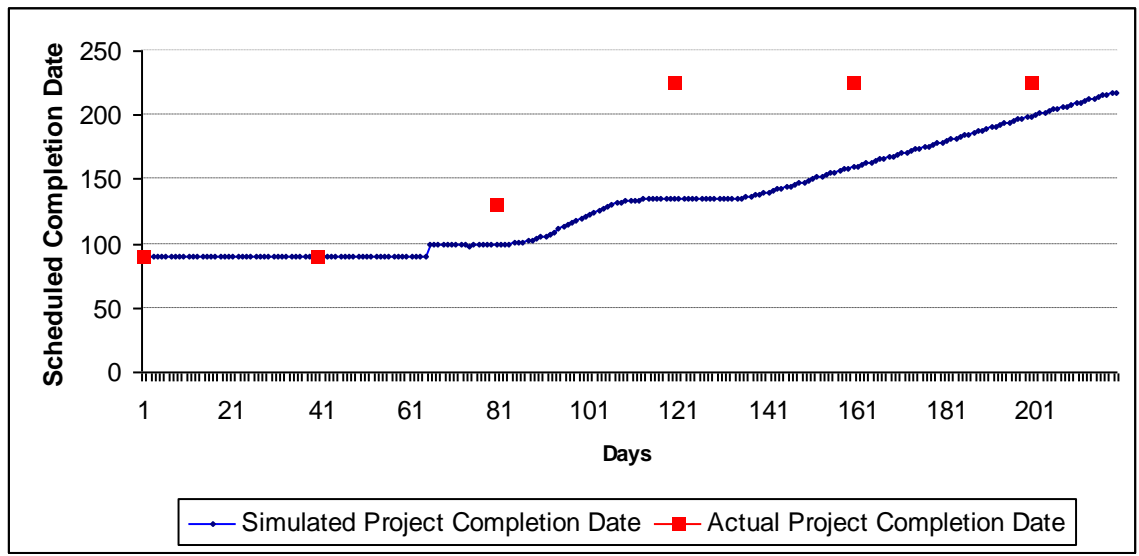

Figure 5. Schedule completion date

The model portrays variation of the expected date of completion of the simulated project activities pertaining to project development, quality assurance and rework, and testing and correction. A continuous adjustment could be noted in response to the on-going status. Initially with change order generation happening at a nearly uniform rate, the project is perceived to on schedule and hence no adjustment is made to the project completion date. With progress, delays are perceived in project status. Coupled with that, the temporal increase in change order generation also resulted in an increase in schedule pressure (not shown). This higher workload necessitated elongation of project schedule which is adjusted accordingly. The simulation output indicated the final completion date of the above mentioned project activities at 218 working days.

In the actual project, adjustments to the schedule were made twice in negotiation with the business user representatives. The first adjustment was made after about four months (80 working days) from the start of project development where the unit testing was postponed by about two-and-half months (50 working days). The final adjustment to project schedule took place when about 75% of the added working days have been expended. The project organization faced situations when some features had to be included in the planned release, and this was only possible by extending the completion date. The actual completion date of the project as derived from the project metrics came out as 65 calendar weeks (325 working days). This also included requirements analysis & prototyping, offshore infrastructure set-up, implementation & acceptance testing and subsequent support phase, which are outside the model boundary. The total estimate of these was found to be 80 working days made at the start of the project. In absence of the actual estimates of these, subtracting the figure leads to equivalent working days of 245, very close to the model outcome. The deviation explained by the absence of related information from the collected project metrics.

**Cumulative Effort Expended**

The simulation model indicated an effort expenditure of 2566 man-days broken up into the following components as given in Table 3. From the table, the testing effort could be observed to be 27% higher than the estimate (736 man-days). The difference is contributed by the elongation of the project's schedule (Figure 5), which meant that the project workforce had to spend more time on the last phases of the project i.e. testing and associated corrections.

Table 3. Effort breakup (simulation output)

| Effort Components | Value |
|---|---|
| Development Effort | 1281 man-days |
| Quality Assurance (QA) Effort | 275 man-days |
| Rework Effort | 10 man-days |
| Training Effort | 65 man-days |
| Testing Effort | 935 man-days |
| Total | 2566 man-days |

The actual effort that was expected on the real project on these activities was found to be 2452 man-days which is close to the simulation results. The slight variation in these two results can be explained based on the differences in the QA effort. In the real project the quality assurance activities were carried out by a fixed number of personnel irrespective of the team size. The simulation model assumed that the QA team size is a fraction of the project manpower, and hence varied with the team size. The resulting difference contributed to the said variation.

**Number of Errors**

Finally, Figure 6 depicts the number of errors generated during the simulation run. The red squares represent the actual number of errors that were committed.

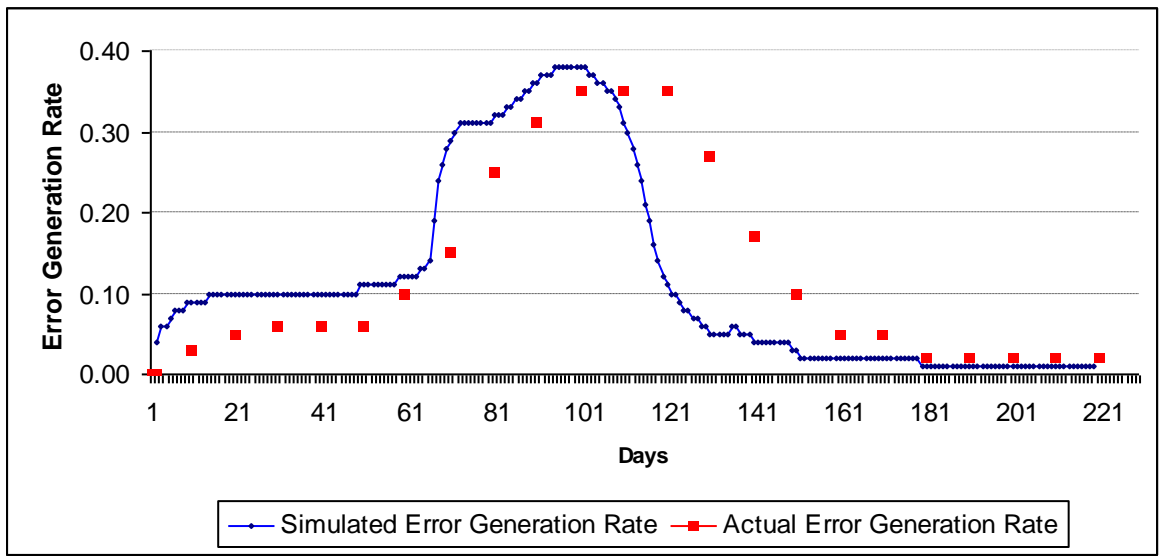

Figure 6. Error generation rate

In the model, the error generation rate is measured as the proportion of the task development rate (number of tasks developed) and the multiplier for quality. The task development rate depends on productivity (as shown in Figure 4) and the workforce committed to development. The multiplier is a factor that combines the impacts of schedule pressure, work force mix and an estimate of the possible number of errors per job size. The pattern of error generation closely follows the software development rate (not shown), which increases after a delay driven by the increase in project workforce size. With time as fewer tasks remains to be processed, the software development rate falls identically affecting the error generation rate (Figure 6).

The actual scenario of error generation is found to resemble the simulated outcome very closely. Most of the errors were caused during the coding stages of the project when the error generation rate increased. A couple of important change requests were also raised in this period which contributed to the effect. In the later stages, as most of the work was already accomplished the error rate died down.

## 5. CONCLUSIONS AND FUTURE RESEARCH

The objective of the case study is to understand and explain the dynamics that influence software project development under uncontrolled change order generation. The project dynamics model of Abdel-Hamid and Madnick [6] was used, and the model parameters were calibrated to the real-project environment. Results indicated how change order generation following a nearly uniform pattern influenced project performance. Both schedule and effort overrun could be noticed which also contributed to an increase in error generation. The results are also in accordance with a related study [9], where the uniform change order generation rate contributed to maximum effort and schedule overruns. The result also facilitates making the following observation: despite the differences between the simulated and the actual project workforce pattern, the total effort expenditure was found to be extremely close (4.6% variation). In the project there were many instances where the project workforce under management pressure had to work for six or seven days a week. Since the calculations in our simulation model is based on a fixed five day per week working mode, the higher man-day per day effort because of larger project workforce in this case is somewhat balanced by the extra working days with comparatively reduced workforce under the actual case.

The following limitations are worth mentioning at this point. While the model was quite accurate in reproducing the project's patterns of dynamic behavior, the deviations from actual values of the variables were caused by the following important differences between the model structure and the project environment:
- The model doesn't capture the holdup event that happened in the real project thereby disrupting the usual flow of work
- The model overestimates the workforce level. The workforce augmentation took place at discrete intervals in whole numbers in the real project, but the model allows for even fractional changes and following a continuous curve.
- The incorporation of change requests also happened as discrete events in the project, but in the model they vary continuously, leading to changes in project progress rates between the real scenario and the simulated output.
- The initial effort allocation policy in the model is a function of the project size, and thus varies accordingly.

Suitable extensions of this work could be to investigate through simulation the different management policies that could lead to improvement in project performance, and investigating their feasibility in a real project environment. Additionally, multiple case studies can also be conducted in order to analyze how different project environment influences the overall dynamics. In would be interesting to see how this method influences design of change management strategies in organizations.

**Authors**

Rahul Thakurta is an Assistant Professor of Information Systems at Xavier Institute of Management Bhubaneswar, India. His primary research interests are software process and project management, and technology adoption and diffusion. He is also the Managing Editor of Research World, and holder of the DAAD Research Fellowship.

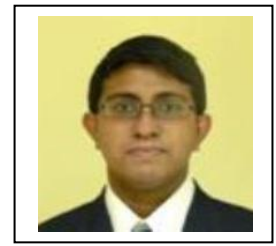

Subhajit Dasgupta is a Senior Technical Consultant at Wipro Technologies and has spent more than a decade on Application development/Maintenance, Implementation and Production Support covering domains like BFSI, Retail and Supply Chain.

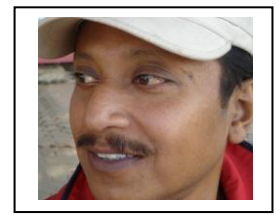